# Microstructural features and hydrogen diffusion in bcc FeCr alloys: a comparison between the Kelvin probe- and nanohardness based-methods

Jing Rao[a], Binhan Sun[a,b], Arulkumar Ganapathi[a], Xizhen Dong[a], Anton Hohenwarter[c], Chun-Hung Wu[a], Michael Rohwerder[a], Gerhard Dehm[a *], Maria Jazmin Duarte[a *]

[a]Max-Planck-Institut für Eisenforschung, Max-Planck-Straße 1, 40237 Düsseldorf, Germany

[b]Now at State Key Laboratory of Chemical Safety, East China University of Science and Technology, Shanghai 200237, China

[c]Department of Materials Science, Montanuniversität Leoben, Jahnstraße 12, 8700 Leoben, Austria

*corresponding authors: Maria Jazmin Duarte j.duarte@mpie.de, Gerhard Dehm dehm@mpie.de

**Abstract**

Hydrogen embrittlement can result in a sudden failure in metallic materials, which is particularly harmful in industrially relevant alloys, such as steels. A more comprehensive understanding of hydrogen interactions with microstructural features is critical for preventing hydrogen-induced damage and promoting a hydrogen-based environment-benign economy. We use the Kelvin probe-based potentiometric hydrogen electrode method and thermal desorption spectroscopy to investigate hydrogen interactions with different hydrogen traps in ferritic FeCr alloys with different chromium contents, dislocation densities, and grain sizes. In addition, we confirm the validity of a novel nanohardness-based diffusion coefficient approach by performing *in situ* nanoindentation testing. Simultaneous acquisition of the dynamic time-resolved mechanical response of FeCr alloys to hydrogen and the hydrogen diffusivities in these alloys is possible during continuous hydrogen supply. Dislocations, grain boundaries and Cr atoms induce reversible hydrogen trapping sites in these ferritic alloys, leading to the reduction of the hydrogen diffusion coefficients and the increase of the absorbed hydrogen.

**1. Introduction**

Hydrogen shows great promise as a future energy resource with no carbon dioxide emissions, as its only by-product is water. The European Green Deal aims to reduce greenhouse gas emissions by 55% by 2030, further highlighting the importance of hydrogen in achieving these environmental targets [1]. However, a severe problem known as hydrogen embrittlement (HE) hinders the application of metals in the hydrogen economy [2]. Hydrogen prefers to accumulate around the stress-concentrated regions and has a high diffusivity of up to $10^{-5}$ to $10^{-4}$ $cm^2/s$ in ferritic iron alloys even at ambient temperature [3, 4]. Due to its small atomic radius, hydrogen can easily be absorbed and diffused through the material. Even with a low hydrogen solubility of less than 3 at. ppm at room temperature, pure iron still faces the issue of HE [5]. Therefore, understanding the diffusion behavior of hydrogen and its interaction with different microstructural features in ferritic alloys are critical in unraveling the underlying mechanisms that lead to HE. This knowledge supports producing HE resistant metals, therefore preventing their sudden early failure.

When introduced into the metal, interstitial hydrogen atoms can either diffuse through the lattice or be attracted to various types of crystalline defects. These defects also interact with the stress fields generated by hydrogen, acting as different types of hydrogen traps. Based on the hydrogen activation energy ($E_a$), the hydrogen trapping sites are classified into two categories: flat trapping sites (also called reversible hydrogen traps) and deep trapping sites (known as irreversible hydrogen traps). Typically, reversible hydrogen trapping sites have activation energies of less than 50 kJ/mol, differing from the irreversible hydrogen trapping sites that demand activation energies higher than 50 kJ/mol at ambient temperature. This disparity in activation energies effectively hinders the movement of hydrogen within the material, making it more challenging for hydrogen to escape from irreversible trapping sites compared to the reversible ones [3, 6-9]. At specific service temperatures, hydrogen can diffuse almost entirely out of the reversible trap site with lower activation energy after a certain time, while external thermal energy is required to release deeply trapped hydrogen [10]. The reversible trapping sites and the diffusive hydrogen play a more critical role in HE at ambient temperature as the deeply trapped hydrogen tends to lack diffusivity and contribute less to the damage evolution [11].

As generally elucidated in the literature, hydrogen interstitials can be segregated or trapped at grain boundaries [12-15]. According to the hydrogen-enhanced decohesion (HEDE) mechanism, the cohesive strength of the interfaces is significantly reduced by the accumulated hydrogen [16]. Therefore, investigating hydrogen diffusion behavior around the grain boundary region is crucial. Controversial arguments about the hydrogen affinity to the grain boundaries with different characters have been addressed. They can act as reversible hydrogen trapping sites like the Σ3 twist grain boundary or as deep hydrogen trapping sites. Based on the tight-binding model, the Σ9, Σ11 and Σ17 boundaries perform as deep hydrogen traps in α-Fe at room temperature according to the simulations of Eunan et al. [17]. In polycrystalline nickel, Oudriss et al. [18] recognized a hydrogen trapping behavior around "special" $Σ3^n$ grain boundaries by performing secondary ion mass spectrometry in combination with electron backscatter diffraction (EBSD) analysis of the microstructure. On the contrary, some works declare that grain boundaries may act as a short diffusion pathway that stores little hydrogen depending on the atomic arrangement of the grain boundary atoms [12, 19, 20].

Dislocations are critical for the plastic deformation of metallic materials. The well-established hydrogen-enhanced localized plasticity (HELP) mechanism treats dislocations as the trigger for the HE as hydrogen shields the dislocations, enhancing the dislocation activities [21]. Besides, Kirchheim proposed the "defactant theory" based on thermodynamic calculations, explaining the easier dislocation nucleation caused by a hydrogen-induced reduction of the dislocation line energy [22]. In addition, the dislocation-grain boundary interaction is a potential cause for hydrogen-induced intergranular fracture as the impingement of the dislocations pile-up in the vicinity of the grain boundary might be facilitated, as proposed by Novak et al. [23]. Therefore, a comprehensive understanding of hydrogen-dislocations interaction is crucial in unraveling HE mechanisms. This hydrogen impact on dislocation activities has been investigated through electron channeling contrast imaging (ECCI), transmission electron microscopy (TEM) and numerical calculations [24-29]. Indirect observations of deuterium trapping at dislocations and grain boundaries in bcc martensitic steel were also achieved by cryogenic atom probe tomography (APT) [30]. From a macroscopic and quantitative perspective, it is noticed that dislocations and vacancies are the main trapping sites observed by scanning Scanning Kelvin probe force microscopy (SKPFM) and thermal desorption spectroscopy (TDS) in bcc Fe-5 wt.% Ni [31]. Furthermore, by applying a quadrupole mass spectrometer, Ono and Meshii [15] concluded that dislocations are the leading hydrogen trapping site in single phase alpha iron.

As a principal substitutional element, Cr is commonly incorporated into steels to enhance their corrosion resistance and to mitigate neutron irradiation-induced expansion in industrial applications [32]. These applications often entail exposure to harsh environments that carry a risk of HE. The substitutional solute Cr atoms in ferritic steels are generally treated as the cause for the appearance of reversible hydrogen trapping sites. As calculated by simulations, the binding energy for Cr atoms with hydrogen in ferritic iron is 26-27 kJ/mol [33]. Cr then affects the hydrogen diffusion behavior following the chemical and elastic interactions between the Cr atoms and hydrogen. Regarding chemical interactions, Cr has a higher chemical (electronically) affinity to hydrogen than Fe based on the Pauling scale [34]. While considering the elastic interaction, the Cr atom has a larger atomic radius than a Fe atom, generating a distortion around the Cr atoms within the Fe matrix. A reduction of diffusion coefficient with higher Cr content due to the retardance of the strain field around Cr atoms was commonly observed by permeation tests. Hagi found this effect in Fe-0.4~5.1 at.%Cr by the electrochemical permeation technique [35]. The interstitial sites formed around substitutional atoms are believed to serve as hydrogen trapping sites, primarily due to the alteration of strain fields, based on investigations of lattice constant changes using the X-ray diffraction technique [35]. The same reduction in hydrogen diffusion coefficient and enhancement in hydrogen solubility with increased Cr content in FeCr binary alloys was monitored at ambient temperature by the electrochemical double cell and hot extraction chemical analysis, respectively [3]. Even at a higher temperature range of 473-573 K, the reduction in diffusion coefficient among Fe-4~14 wt.%Cr has also been reported using the gas evolution permeation technique [36]. Furthermore, as proposed by Ramunni based on ab-initio simulations, the mismatch originating from Cr atoms produces a more stable octahedral interstitial site for hydrogen that reduces the diffusion coefficient by two orders of magnitude and enhances the hydrogen solubility in Fe-9 wt.%Cr compared to annealed α-Fe [37]. On the other hand, Cr carbides are generally treated as irreversible hydrogen trapping sites in ferritic steels [27, 33].

Different analytical methods have been used to characterize the trapping of hydrogen at different microstructural features, such as small-angle neutron scattering [38], transmission electron microscopy [39] and atom probe tomography [30]. Although these methods detect hydrogen/deuterium with high spatial and/or chemical resolution at a specific location, there is a lack of quantitative information on diffusive hydrogen at a macroscopic level, which prevents the applicability of microstructural engineering from effectively mitigating HE effects. TDS is critical in investigating the thermodynamic and kinetic behavior of desorption/decomposition procedures in bulk materials. The amount of hydrogen desorbed from various hydrogen trapping sites at a defined heating ramp can be quantitatively identified by this technique as it is microstructural constituent sensitive [31].

However, the loss of diffusible hydrogen prior to the TDS measurement is inevitable and the hydrogen diffusion rate throughout the specimen is unable to be determined. As compensation for thermal destructive TDS measurements, alternative permeation techniques are commonly applied where the hydrogen is continuously generated and penetrates through the specimens. The Devanathan-Stachurski (DS) double cell, based on the principle theory of Fick's law, is widely adopted in conventional hydrogen permeation studies, enabling a quantitative understanding of the hydrogen diffusion coefficient by measuring the hydrogen oxidation current at the exit side of the second electrochemical cell [40]. Nonetheless, the palladium layer deposited on the sample surface using the physical vapor deposition (PVD) method exhibits instability under harsh polarization conditions as it is immersed into the electrolyte. Therefore, in this study, a non-destructive Kelvin probe (KP)-based potentiometric technique was applied, enabling the measurement of localized potentials, with high hydrogen detection sensitivity instead of the averaged permeation rate in DS double cell [41-45].

Investigating the hydrogen-induced degradation of the material's mechanical properties is vital in studying the HE mechanisms. *In situ* hydrogen charging techniques are ideal for exploring the response of ferritic materials to hydrogen due to their high hydrogen diffusivity [46, 47]. Special interests have been triggered in the *in situ* nanoindentation method brought up by Barnoush and Vehoff [46], which can introduce localized deformation into the material with a lateral resolution in the nanometer-scale while simultaneously charging the specimens electrochemically with hydrogen. However, the possible electrolyte-induced surface degradation [48, 49], stimulated the development of a novel backside hydrogen charging method [47, 50]. The chemistry and surface topography remain unchanged on the testing (front) side of the specimen during hydrogen charging from the back side, enabling to perform mechanical and permeation investigations in bulk bcc FeCr alloys at the same time.

In this study, we investigate the hydrogen response to different microstructural features, including dislocations, grain boundaries, and Cr content, in model ferritic FeCr alloys. The specified microstructural features were produced through different metallurgical treatments and characterized using EBSD and ECCI. The immobile hydrogen content absorbed by the materials after saturation charging has been quantitatively determined using TDS. An *in situ* KP-based potentiometric hydrogen electrode technique has been applied to measure the hydrogen diffusion coefficients. This method is compared to a new nanohardness-based approach for measuring the hydrogen diffusion coefficient. The nanohardness data is collected using an *in situ* backside hydrogen charging setup, enabling the simultaneous measurement of nanohardness and Young's modulus during electrochemical hydrogen charging [50]. The interaction between hydrogen and various microstructural features can be more comprehensively understood by combining results from TDS with investigations using the KP-based permeation method.

## 2. Experimental procedures

### 2.1. Materials

Table 1 displays the principal chemical composition of the model ferritic alloys Fe-9 at.% Cr, Fe-16 at.% Cr, and Fe-21 at.% Cr (denoted as Fe-9Cr, Fe-16Cr, and Fe-21Cr hereafter) that were fabricated in-house. The necessary raw elements were weighted, melted, cast, hot rolled and annealed under Ar protection atmosphere. A homogeneous microstructure with low dislocation density (LD) and a large grain (LG) size in the millimeter range was obtained as described in [50]. Fe-21Cr with high dislocation densities (HD) were produced by cold rolling at ambient temperature to achieve thickness reductions of 5 % (named as CR5) and 10 % (named as CR10). Smaller grain (SG) sized specimens were processed by the high-pressure torsion (HPT) technique, details see [51]. The following heat treatment for 10 min at 600 °C was utilized to reduce the dislocation density of the material.

Table 1      Main chemical composition of the materials investigated (in at.%).

| Element | Fe | Cr | C | O |
|---|---|---|---|---|
| Fe-9Cr | Balance | 8.60 | 0.08 | 0.09 |
| Fe-16Cr | Balance | 15.92 | 0.01 | 0.07 |
| Fe-21Cr | Balance | 20.94 | 0.01 | 0.04 |

*2.2. Microstructure characterization*

The dislocation density was characterized using a field-emission scanning electron microscope (SEM, Carl Zeiss AG, Zeiss-Merlin SEM) with the electron channeling contrast imaging (ECCI) method under two beam conditions. This technique possesses high spatial resolution for bulk materials to distinguish individual dislocations, as well as high statistical reliability with a sufficiently large inspection zone [52]. The applied beam current was 2 nA with an accelerating voltage of 30 kV and a working distance of ~7.5 mm. Dislocations were quantified by the software Fiji [53]. At least three randomly chosen regions (~100×100 $\mu m^2$) were used in LG samples to calculate dislocation densities and more than 3 grains were selected in SG specimens. Electron backscatter diffraction (EBSD, Jeol, JSM 6490 SEM) was used to obtain the crystallographic orientation of each grain, as well as quantification of the grain size based on the area including more than ~40 grains with the help of TSL OIM software.

*2.3. Nanohardness-based in situ electrochemical hydrogen permeation experiment*

The nanohardness-based hydrogen diffusion coefficients in Fe-16Cr and Fe-21Cr alloys (LG, LD) were measured by the in-house *in situ* electrochemical backside hydrogen charging setup assembled inside the nanoindentation chamber (Fig. 1a).

Details about the setup were described in [50]. The specimens were cut to a disk with a thickness of ~2.4 mm and a diameter of ~13 mm by electrical discharge machining to reduce the thickness of the deformation layer. The surface was mechanically grounded consecutively with 1000-4000 grit SiC papers, followed by wet polishing with diamond suspensions untill 1 µm. The final surface treatment was then applied, combining chemical etching with agent $V_2A$ for 10 s and mechanical polishing with 40 nm colloidal silica suspension for 1 h. After freshly polishing to avoid the influence of the native oxide, the final surface roughness obtained was less than 1 nm/µm$^2$ as analyzed by atomic force microscopy (AFM, Dimension$^{TM}$ 3000).

Hydrogen was produced in a three-electrode cell within a hydrogel electrolyte that contains 0.1 M NaOH and an addition of 20 mg/L $As_2O_3$ as the hydrogenation promoter, for the generation of monatomic hydrogen instead of diatomic hydrogen [54]. The backside of the specimen that contacts the electrolyte is a circular region of 50.27 mm$^2$. All the electrochemical potentials were collected with respect to the commercial Ag/AgCl reference electrode (RE, World Precision Instruments, DRIREF-2SH) as $V_{Ref}$. The applied potentials were chosen from the carried out cyclic sweep voltammetry curve to reach a constant current between -0.5 to -2 mA/cm$^2$, producing a considerable amount of hydrogen through the hydrogen evolution reaction (HER) while inhibiting the instability introduced by the excess hydrogen bubbles.

During hydrogen charging, nanoindentation was conducted in a load-controlled instrument (KLA G200, former Agilent) for a maximum load of 10 mN with a constant strain rate of 0.1 s$^{-1}$. The Polychlorotrifluoroethylene (PCTFE) manufactured setup body has a low water absorption rate and high stiffness, therefore withstands the electrolyte corrosion and enables the mechanical measurements reliably performed by nanoindentation. Details about the electrochemical hydrogen charging setup are given in [50]. An Argon flux (3.5 L/min) purged through a pipe above the sample's surface was utilized to minimize the loss of hydrogen by reducing the oxygen content.

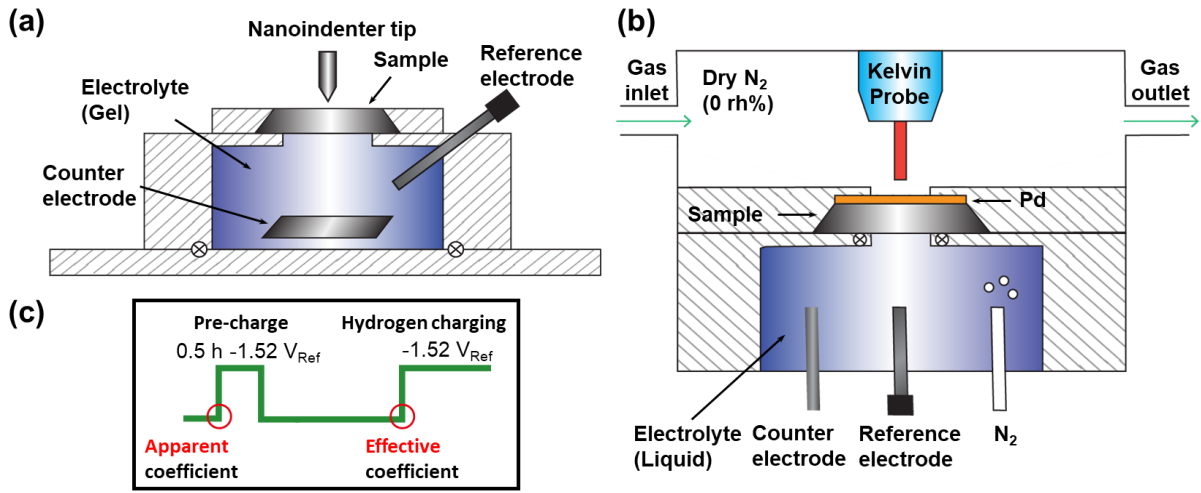

Fig. 1. (a) A schematic diagram of the *in situ* backside hydrogen electrochemical charging nanoindentation setup (reproduced from [50]). (b) The sketch of the hydrogen electrode-based hydrogen permeation measurement setup (reproduced from [55]). (c) The sketch of the KP-based potentiometric hydrogen electrode experimental procedure. The apparent diffusion coefficient is obtained before the hydrogen pre-charging, while the effective diffusion is acquired after the hydrogen pre-charging.

Nanohardness was obtained by the continuous stiffness measurement (CSM) method based on Oliver and Pharr's theory with a standard Berkovich diamond tip of radius (~400 nm) as calculated from the Hertzian equation [56-59]. The elastic modulus and hardness values were calculated using a Poisson's ratio of 0.3, as commonly applied for steels, between the indentation depth of 250-350 nm, considering the indentation size effect [60]. As an interpretation of environmental thermal and mechanical vibration for the setup, the thermal drift rate was kept below 0.05 nm/s during the nanoindentation tests. In addition, A 60s holding at 10% of peak load during unloading measured the thermal drift rate, allowing accurate adjustments to the load-displacement curve [61].

After filling up the setup with the hydrogel electrolyte, the influence of the deeply trapped hydrogen was avoided by pre-charging the samples for 3 h with a current density of 2 mA/cm$^2$, followed by the release of the lightly trapped hydrogen and diffusive hydrogen for 6 h. Afterward, the reference hardness and the elastic modulus were collected. The recharge of Fe-16Cr and Fe-21Cr was then commenced with a potential of -1.25 $V_{Ref}$ (~1.3 mA/cm$^2$), aiming to probe the evolution of the nanohardness and the hydrogen diffusion coefficient calculation.

*2.4. Thermal desorption spectroscopy (TDS) tests*

The quantification of the hydrogen content in the different materials was conducted by TDS measurements (Hiden TPD Workstation). Both sides of circular specimens with a diameter of 10 mm and a thickness of 1 mm were ground with 600-2000 grit SiC papers to eliminate the oxide layer. Afterward, the thickness of the specimen was captured three times to ensure accuracy by optical microscopy, avoiding contact with the sample surface. Electrochemically charging the specimens with hydrogen for 4 h with a constant current density of ~2 mA/cm$^2$ is sufficient for thoroughly filling the specimens with hydrogen, considering the hydrogen diffusion coefficient. Afterward, the specimens were immediately removed from the electrolyte composed of 3 wt.% NaCl and an addition of 0.3 wt.% $NH_4SCN$ to poison the hydrogen recombination process [62]. Before putting the H-charged sample into the TDS chamber, they were cleaned with deionized water and acetone to remove the remaining electrolyte. Within 20 min, the TDS measurement was initiated with a heating rate of 16 °C/min from ambient temperature to 470 °C.

*2.5. Kelvin probe-based potentiometric hydrogen electrode experiments*

The hydrogen diffusion coefficients in different materials were obtained from a Kelvin probe-based (KP-based) potentiometric hydrogen permeation experiment, introduced by Evers et al. [42]. The setup is composed of 2 parts (Fig. 1b): 1) the lower part is the KP microscope (Wicinski&Wicinski Surface Scanning Systems) that investigates the potential variation of the hydrogen electrode on the Pd-coated side that is exposed to a continuous dry nitrogen flux (0 rh %) [43]; 2) the upper part serves to continuously charge hydrogen electrochemically into the samples.

This study investigated two kinds of hydrogen diffusion coefficients, the apparent diffusion coefficient ($D_{app}$) and the effective diffusion coefficient ($D_{eff}$), as exhibited in Fig. 1c. The $D_{app}$ includes the effects from irreversible hydrogen, reversible hydrogen and interstitial diffusive hydrogen, while $D_{eff}$ is attained after filling the deep hydrogen traps and considers only mobile and lightly trapped hydrogen. Samples for measuring the $D_{eff}$ have the same dimension as that used in the *in situ* electrochemical backside hydrogen charging tests. For the HPT-treated samples, the operation duration (time lag for the H permeation) was drastically enhanced by introducing grain boundaries, i.e. trap sites, resulting in related problems with the experimental system and leakage of the electrolyte during the required long durations of the corresponding experiments. Hence, the thickness of the samples was reduced by half to ~1.1 mm. Both sides of the samples were ground with 600-4000 grit SiC papers to remove the oxide layer. Before performing the physical vapor deposition (PVD, Leybold Univex 450) of 100 nm Pd thin film on the KP-measured side, the samples were prepared by rinsing with ethanol and $CCl_4$ to remove the organic remains. The Ni80Cr20 KP tip was manufactured by electropolishing until reaching the diameter of 100 µm [42, 44]. The same electrolyte as *in situ* backside electrochemically nanoindentation measurement (0.1M NaOH + 20 mg/L $As_2O_3$ + deionized water) was applied. Pt foil (99.99 %, Goodfellow) was applied as the counter electrode (CE). The standard Ag/AgCl (12.5 cm, Metrohm) was applied as the reference electrode (RE). The current density of ~2 $mA/cm^2$ was designated by setting the constant potential of -1.52 $V_{Ref}$ to the specimens, considering the area of the electrolyte contact region (12.57 $mm^2$ calculated by the sealing O-ring with a diameter of 4 mm). More details of the electrochemical hydrogen charging parameters related to the KP-based potentiometric setup were described in [44].

The reference potential was continuously recorded by the KP microscope. The filling of the electrolyte and the pre-charging were executed simultaneously (-1.52 $V_{Ref}$) until the measured KP potential reached a plateau. The hydrogen-induced potential variation is confined to a certain range of 0.3-0.6 V, depending on the surface condition and minor preexistence of hydrogen before charging. An introduction of above ~2.4 at.% hydrogen into the Pd at ambient temperature initiates the formation of β-H-Pd, a palladium hydride [42]. The potential measured at the state with the coexistence of α-Pd-H (phase with dissolved hydrogen) and β-H-Pd is treated as the reference potential for the calibration. After several trials, the filling duration of the deep traps through pre-charging of the sample was set to 0.5 h before the system reached half of the saturation potential. This reserves a potential volume for the following hydrogen charging step to collect enough data for apparent diffusion coefficient analyses. Once the potential became stable, hydrogen charging was reapplied with the same voltage of -1.52 $V_{Ref}$ to measure $D_{eff}$ exempt from the influence of deep hydrogen traps.

# 3. Results and interpretation

## 3.1. Characterization of microstructure

Grain sizes and dislocation densities for the investigated ferritic alloys with different chemical compositions and metallurgical treatments are listed in Table 2. Based on EBSD measurements, the grain size was calculated with the help of TSL OIM software (Fig. 2). The straight vertical lines in between segments of the EBSD images correspond to stitching lines of the combo scanning technique employed to scan regions of a few millimetres. In Fig. 3, ECCI [52] reveals bright appearing dislocation lines, under two beam conditions, for alloys with different further thermomechanical processing (before and after cold rolling, HPT treated combined with annealing).

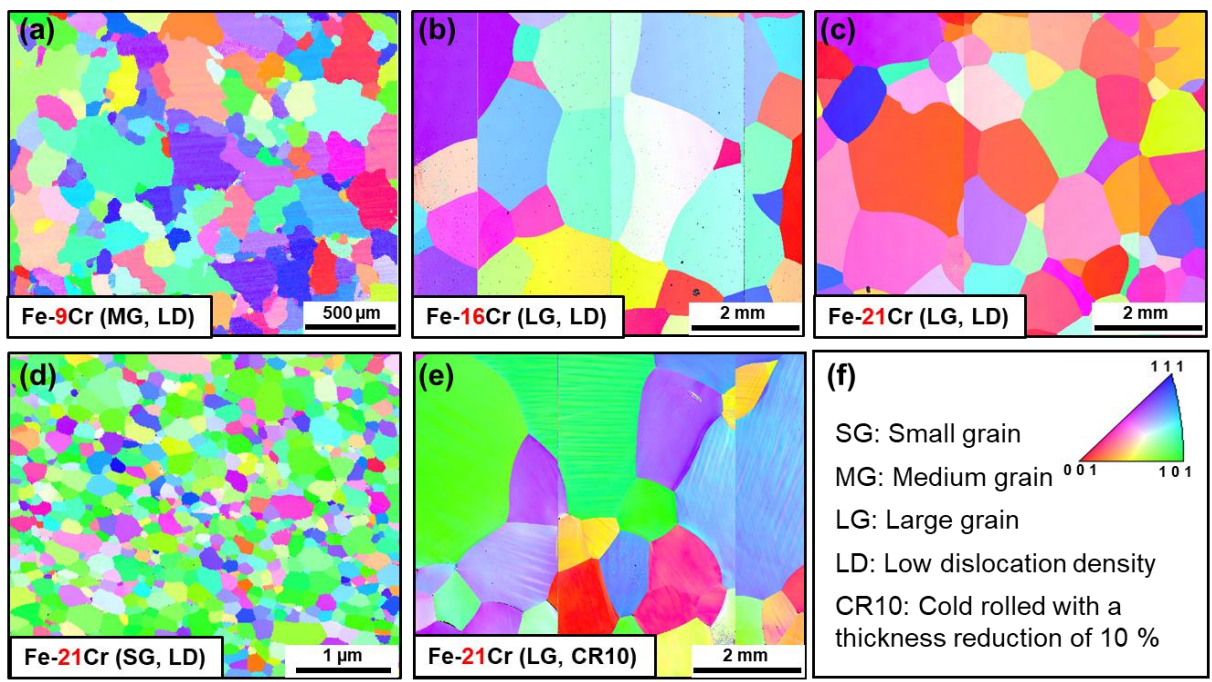

Fig. 2. EBSD images of: (a) Fe-9Cr (MG, LD), (b) Fe-16Cr (LG, LD), (c) Fe-21Cr (LG, LD), (d) Fe-21Cr (SG, LD) and, (e) Fe-21Cr (LG, CR10). The symbols represent the grain size and dislocation density of the alloys as indicated in (f).

The dislocation density was calculated by dividing the number of dislocations by the inspection area. The lowest dislocation density is observed in the base material in Fig. 3a. An intermediate dislocation density was noticed in the material after HPT treatment, followed by annealing as displayed in Fig. 3b, while, after cold rolling, the dislocation density was enhanced drastically (Fig. 3c). The calculated grain sizes and dislocation densities for the different alloys are shown in Table 3.

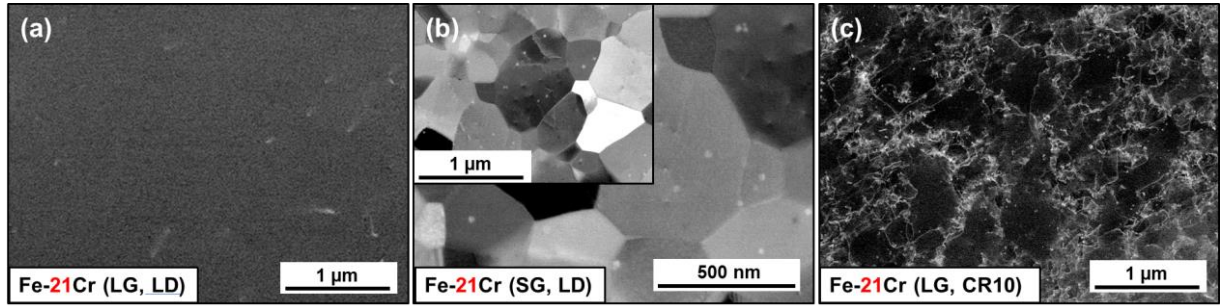

Fig. 3. Dislocation density measurements by ECCI: (a) Fe-21Cr (LG, LD), (b) Fe-21Cr (SG, LD) and, (c) Fe-21Cr (LG, CR10). (SG: Small grain size, LG: Large grain size, LD: Low dislocation density, CR10: Cold rolled with a thickness reduction of 10 %)

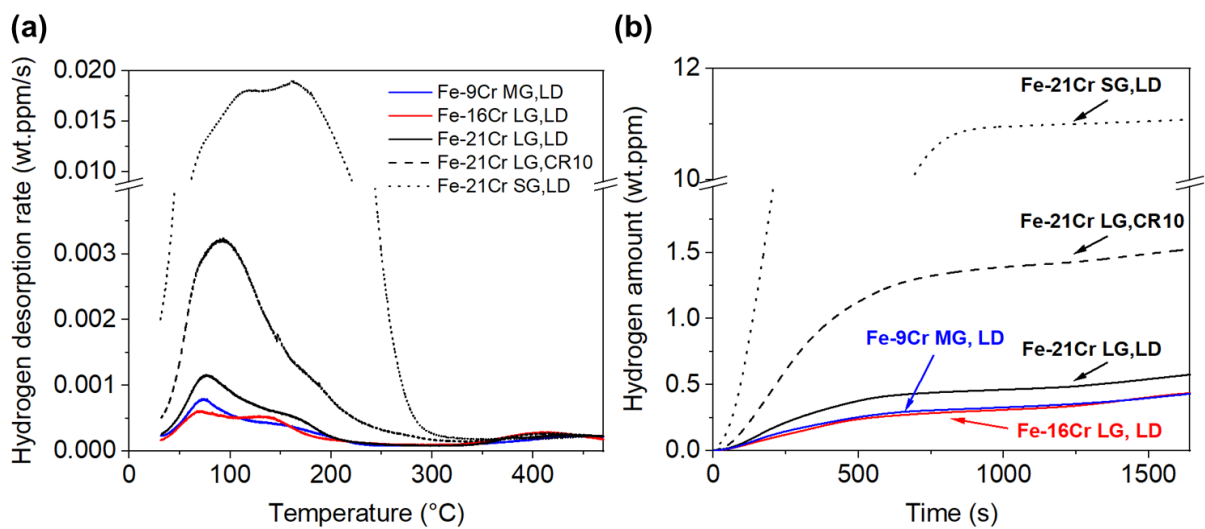

Fig. 4. (a) Hydrogen desorption curves and (b) Hydrogen amount curves for Fe-9Cr (MG, LD), Fe-16Cr (LG, LD), Fe-21Cr (LG, LD), Fe-21Cr (SG, LD) and Fe-21Cr (LG, CR10) after hydrogen charging for 4 h at ~2 mA/cm$^2$. TDS was conducted with a heating rate of 16 °C/min.

*3.2. Hydrogen concentration in different materials investigated by TDS*

TDS investigations were conducted to compare the quantitatively measured hydrogen amount in different treated materials (as annealed, cold-rolled specimens and HPT treated samples), as exhibited in Fig. 4. Fig. 4a shows the hydrogen desorption rate versus temperature curves, the peaks of which indicate different types of hydrogen traps. There exist three peaks of 79±10 °C, 136±18 °C and 435±19 °C as analyzed using the Origin software, with the peaks differentiated based on the Gaussian peak function as shown in Fig. S1.

Table 2    The TDS measured desorbed hydrogen content ($C_{TDS.H}$) and KP-based permeation experiments measured apparent hydrogen solubility ($C_{KP.H}$) in different materials.

|  | Fe-9Cr MG, LD | Fe-16Cr LG, LD | Fe-21Cr LG, LD | Fe-21Cr LG, CR10 | Fe-21Cr SG, LD |
|---|---|---|---|---|---|
| $C_{TDS.H}$ (wt.ppm) | 0.43±0.01 | 0.44±0.02 | 0.58±0.01 | 1.53±0.05 | 11.08±0.12 |
| $C_{KP.H}$ (wt.ppm) | 0.14±0.05 | 0.11±0.03 | 0.15±0.05 | 1.81±0.22 | 11.28±0.53 |

The hydrogen amount $C_{TDS.H}$ (wt.ppm) was determined by calculating the cumulative desorbed hydrogen from the TDS time spectra. By integrating the areas under the peaks, the absorbed hydrogen content in different materials can be obtained as displayed in Fig. 4b. The calculated total amount of absorbed hydrogen is listed in Table 2. The additional values corresponding to $C_{KP.H}$ (wt.ppm) were explained afterwards in the section 4.2 of this manuscript. The Fe-21Cr alloys with their higher Cr content contain a relatively higher amount of hydrogen (0.58±0.01 wt.ppm) compared to Fe-16Cr (0.44±0.02 wt.ppm) and Fe-9Cr (0.43±0.01 wt.ppm). This coincides well with literature [3] that more hydrogen is absorbed in Fe-10.4 wt.%Cr than pure iron after the same hydrogen charging procedure as investigated by hot extraction chemical analysis. When the dislocation density increased from as-annealed of $(2±2)×10^{12}$ m$^{-2}$ to cold-rolled of $(1.4±0.1)×10^{14}$ m$^{-2}$, a ~2.6 times enhancement of absorbed hydrogen in Fe-21Cr is observed as dislocations are suggested to be the hydrogen reversible trapping sites by TDS analysis [31]. Regarding the grain size, decreasing from a bigger grain size of 1133±85 μm to a smaller grain size of 297±138 nm causes an increase of absorbed hydrogen content from 0.58±0.01 wt.ppm to 11.08±0.12 wt.ppm, indicating the critical role of grain boundaries in storing hydrogen.

*3.3. Analysis of hydrogen permeation curves (Apparent diffusion coefficient)*

In our case, the KP evaluates the surface chemical potential variation of hydrogen electrodes generated in the coated Pd layer without the influence of oxygen. Fig. 5a displays the typical evolution of KP measured electrode potential in the Pd coated side starting from the onset of the hydrogen charging in Fe-21Cr (LG, CR10) until the Pd reaches the binary phase state. The binary phase /reference potential ($E_{bp}$), which represents the state when dissolved α-Pd-H coexists with the β-H-Pd, is set to 0 V for calibration [42]. With the enhancement of hydrogen content in Pd, the Kelvin probe measured potential drops drastically until reaching the lowest point, which stays static once the hydrogen content in Pd exceeds ~2.4 at.%. The monitored KP potential ($E$) can be converted to hydrogen concentration in Pd ($c(H_{Pd})$) based on the Nernst equation [44]:

$$E = E_{SHE}^{*'} + m \cdot \ln(c(H_{Pd})) \tag{1}$$

where $E_{SHE}^{*'}$ represents the standard hydrogen electrode potential, the applied $m$ as measured by Evers et al. [42, 43] is -130 mV/decade for this specific nanocrystalline Pd coating with a thickness of 100 nm (deviation should be noted for Pd layer with different microstructure features). Fig. 5b exhibits the typical permeation curve that shows the corresponding evolution of hydrogen concentration in Pd calculated from Fig. 5a. The curve follows the 3 stages of initial trap filling, steady hydrogen permeation and saturation (also known as binary phase state). The retention of hydrogen penetrating through the specimen is considered as the 'time-lag' [3], which implies the behavior of hydrogen traps within the materials. By measuring the time where the asymptotic linear portion of the rising permeation curve intersects with the x-axis (see Fig. 5b), the lag time ($t_{lag}$) is extracted and further utilized to calculate the hydrogen apparent diffusion coefficient ($D_{app}$) and effective diffusion coefficient ($D_{eff}$) according to the following equation:

$$D = \frac{L^2}{6 * t_{lag}} \tag{2}$$

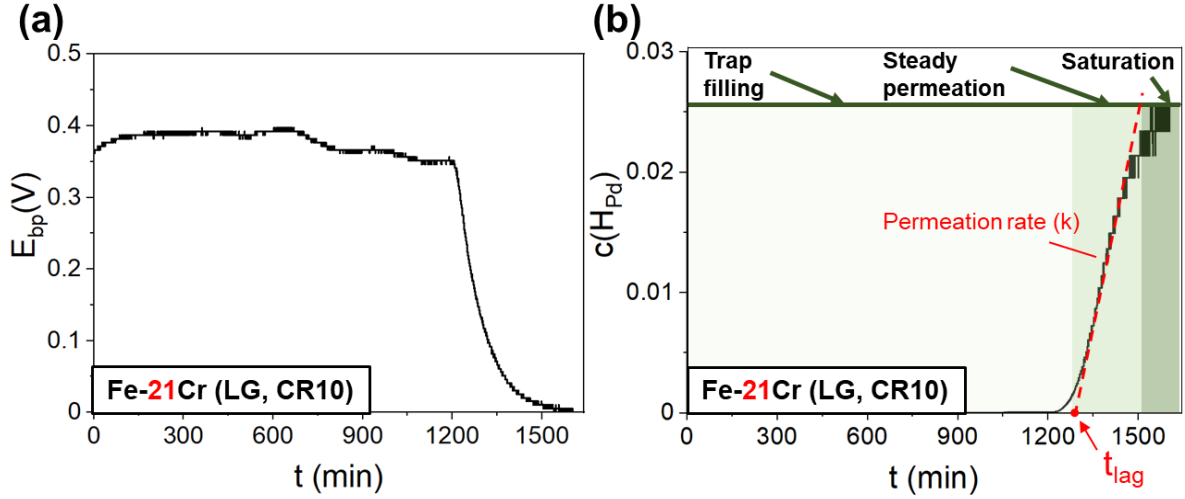

Fig. 5. (a) Evolution of the Kelvin probe potential measured on the 100 nm Pd coated side during hydrogen charging in Fe-21Cr (LG, CR10) at a polarization voltage of -1.52 $V_{Ref}$. (b) Permeation curve transferred based on the Nernst equation, to calculate the apparent hydrogen diffusion coefficient ($D_{app}$) (Eq. (1)), where c($H_{Pd}$) is a measure of the amount of hydrogen that passed through the sample. The slope of the red linear fitting for the permeation curve is considered as the steady state permeation rate ($k$), and its intersection with the x-axis indicates the lag time ($t_{lag}$).

The linear portion of the permeation curve can be fitted and its slope ($k$) can be used to calculated the steady state permeation current ($i_p$) based on the following formula:

$$i_p = F \cdot n_{Pd} \cdot k \tag{3}$$

where F is the Faraday constant, $n_{pd}$ is the molar amount of Pd deposited on top of the sample. A reverse proportional relation is observed between the sample thickness and the permeation current [44].

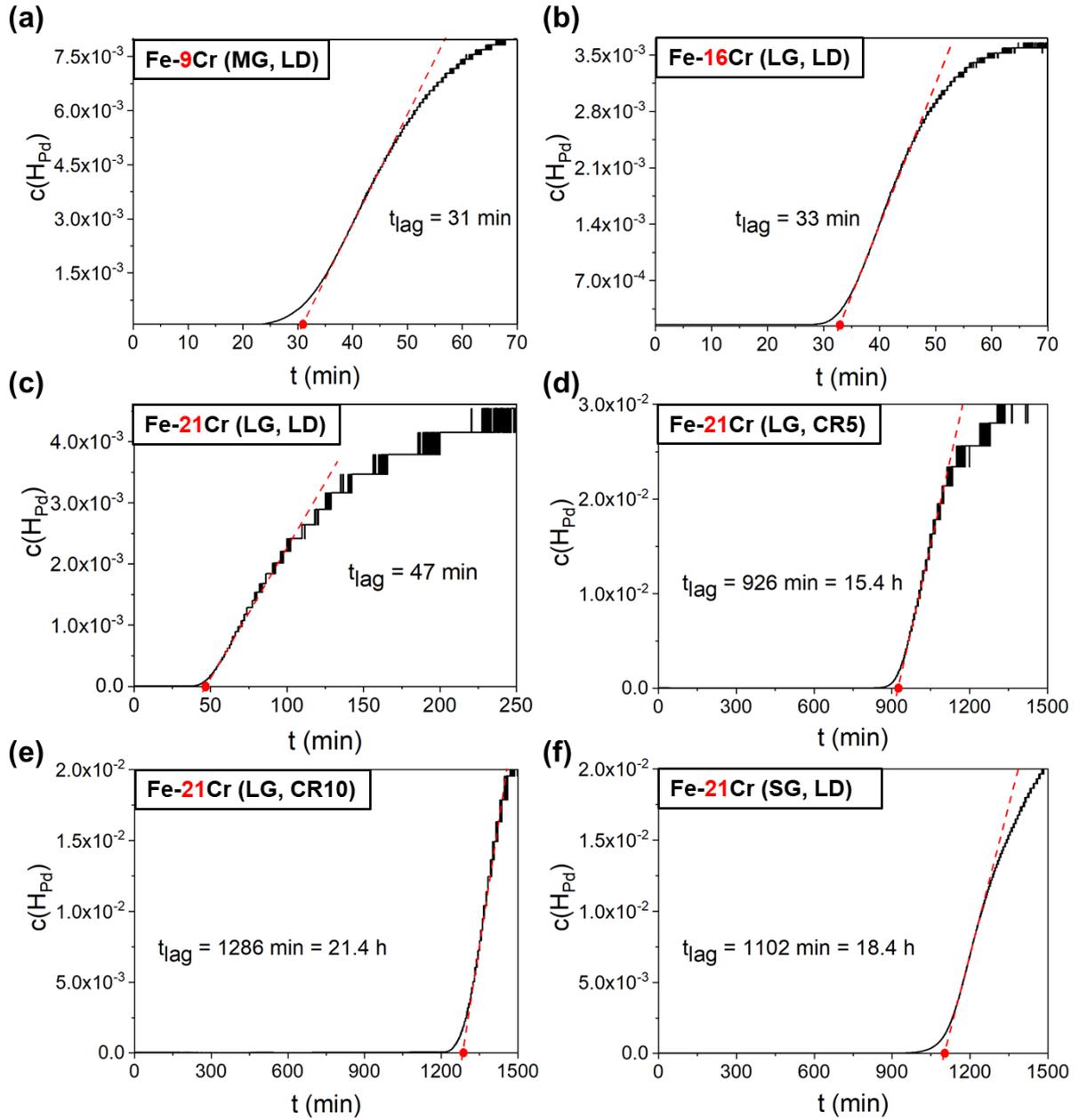

Fig. 6. Permeation curves for the apparent diffusion coefficient calculation based on the KP measured hydrogen electrode potential in the coated 100 nm Pd layer for different materials: (a) Fe-9Cr (MG, LD), (b) Fe-16Cr (LG, LD), (c) Fe-21Cr (LG, LD), (d) Fe-21Cr (LG, CR5), (e) Fe-21Cr (LG, CR10) and, (f) Fe-21Cr (SG, LD). The dotted red line is the slope for the steady-state rate for hydrogen permeation. The intersection of the dotted red line with the x-axis indicates the lag time ($t_{lag}$).

Fig. 6 shows the permeation curves for Fe-Cr alloys with different material conditions of Fe-9Cr, Fe-16Cr and Fe-21Cr (with various dislocation densities and grain sizes) without hydrogen pre-charge for the calculation of the apparent diffusion coefficient.

With the introduction of a continuous hydrogen flux, after filling the deep hydrogen trapping sites and reversible hydrogen trapping sites, the diffusive hydrogen finally reaches the Pd-coated side through interstitial lattice diffusion across the specimen. As opposed to the materials in Fig. 6d-f that are charged to the saturation state, specimens in Fig. 6a-c are charged in accordance with the first step of Fig. 1c, until the system reaches about half its saturation status. As depicted in Fig. 6a-c, Fe-9Cr possesses the shortest lag time of 31 min, whereas Fe-21Cr occupies the longest lag time of 47 min and Fe-16Cr has the intermediate lag time of 33 min under the same hydrogen charging conditions. The lag time is proportional to the number of hydrogen traps in materials, indicating Fe-21Cr has the highest amount of hydrogen trapping sites. A comparison of the permeation behavior between Fe-21Cr with different dislocation densities and grain sizes is shown in Fig. 6c-f. In samples with large grain sizes of the millimeter range, the lag times are prolonged if the dislocation density is large, following the tendency of $t_{lag}(CR10) > t_{lag}(CR5) > t_{lag}(LD)$. Besides, as illustrated in Fig. 4c and f, a longer lag time appears for the Fe-21Cr alloy with a smaller grain size.

In contrast to Fig. 6, the lag time in Fig. 7 was collected after hydrogen pre-charging followed by releasing of both the reversible trapped hydrogen and diffusive hydrogen for Fe-9Cr, Fe-16Cr and Fe-21Cr (Fig. 1c illustrates the procedure). The effective diffusion coefficient can be calculated based on this lag time, considering the delay originated from the reversible traps and hydrogen diffusion through interstitial sites. As shown in Fig. 7 again, an enhancement in the lag time with increasing Cr content is observed. This coincides well with the previous TDS results revealing that alloys with higher Cr content contain more hydrogen, suggesting Cr is acting as a reversible hydrogen trapping site in ferritic Fe-Cr alloys.

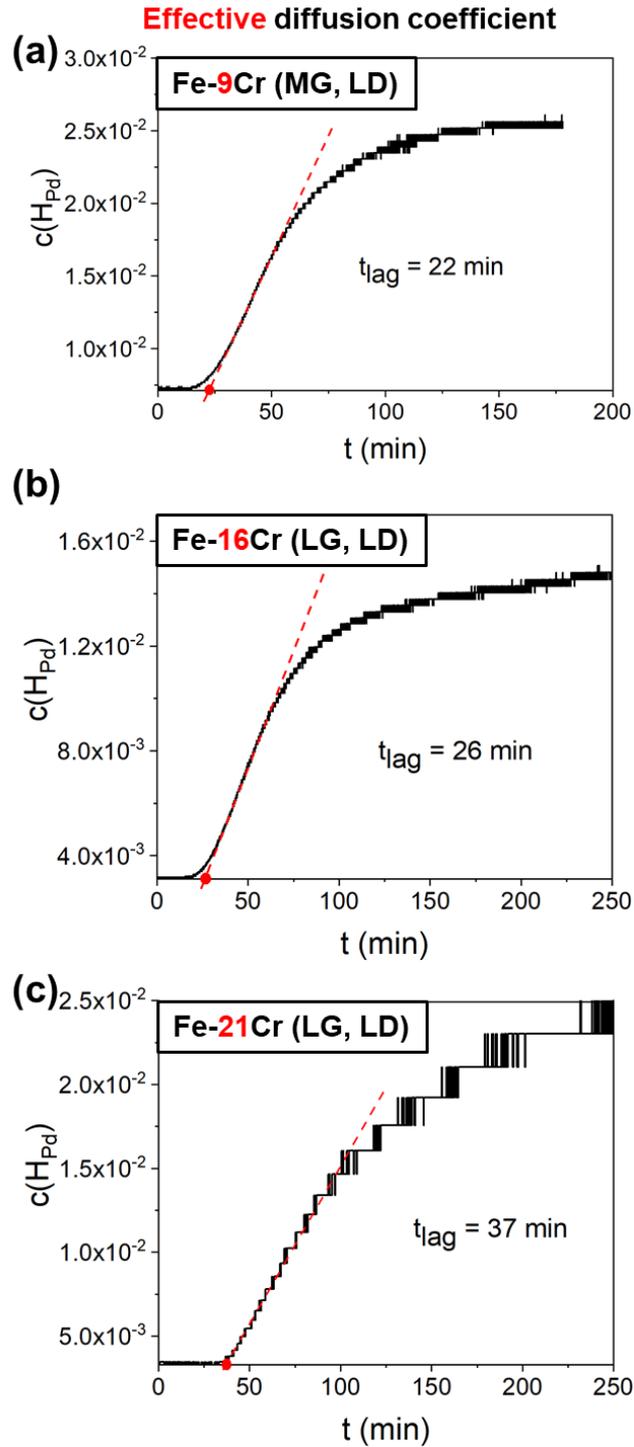

Fig. 7. Permeation curves for the effective diffusion coefficient calculation based on the KP measured hydrogen electrode potential for: (a) Fe-9Cr (MG, LD), (b) Fe-16Cr (LG, LD) and (c) Fe-21Cr (LG, LD).

Table 3 summarizes the apparent hydrogen diffusion coefficient ($D_{app}$) and permeation current ($i_p$) for Fe-9Cr, Fe-16Cr and Fe-21Cr with varying grain sizes and dislocation densities. A reduction in both $D_{app}$ and $i_p$ for samples with higher Cr content, dislocation densities and grain boundaries can be noticed as hydrogen permeation is slower in such samples with those imperfections compared to an ideal defect-free crystal in most cases [63]. Note that the difference in dislocation density between the cold-rolled Fe-21Cr samples of varying thicknesses is small when considering the scale bar. Given such a high dislocation density of $10^{14}$ m$^{-2}$, it is hard to differentiate the subtle dissimilarities using ECCI images. Nonetheless, it remains an efficient technique for investigating materials with different magnitudes of dislocation densities.

Table 3  Overview of grain size, dislocation densities and apparent hydrogen diffusion coefficients $D_{app}$ calculated based on $t_{lag}$ for Fe-9Cr, Fe-16Cr and Fe-21Cr.

| Alloy | Grain size (μm) | Dislocation density (m$^{-2}$) | $D_{app}$ (cm$^2$/s) | $i_p$ (μA/cm$^2$) |
|---|---|---|---|---|
| Fe-9Cr (MG, LD) | 149±33 | (2±2)×10$^{12}$ | (4.40±0.11)×10$^{-6}$ | 0.59±0.02 |
| Fe-16Cr (LG, LD) | 1133±85 | (3±2)×10$^{12}$ | (4.11±0.08)×10$^{-6}$ | 0.33±0.03 |
| Fe-21Cr (LG, LD) | 1049±51 | (2±2)×10$^{12}$ | (2.86±0.03)×10$^{-6}$ | 0.31±0.02 |
| Fe-21Cr (LG, CR5) | 976±109 | (1.3±0.1)×10$^{14}$ | (1.68±0.01)×10$^{-7}$ | 0.21±0.02 |
| Fe-21Cr (LG, CR10) | 959±267 | (1.4±0.1)×10$^{14}$ | (1.29±0.01)×10$^{-7}$ | 0.20±0.01 |
| Fe-21Cr (SG, LD) | 0.3±0.1 | (3.7±0.6)×10$^{13}$ | (1.92±0.12)×10$^{-8}$ | 0.12±0.02 |

*3.4. Mechanical properties obtained before and during in situ hydrogen charging*

Fig. 8 displays the evolution of the Young's modulus (blue) and hardness (red) in Fe-16Cr (LG, LD) and Fe-21Cr (LG, LD) alloys before and during cathodic hydrogen charging by utilizing the *in situ* backside nanoindentation setup. Hydrogen charging is achieved by the potentiostatic method and the corresponding current density at a designated potential is collected and displayed in the green curves. The hydrogen charging procedure is consistent with the second stage of the permeation experiment, as shown in Fig.1c. After pre-charge and release of the lightly trapped and diffusive hydrogen, the mechanical reference data are collected (the red-filled squares). The initial reference hardness of Fe-21Cr is larger than that of Fe-16Cr due to substitutional solid solution strengthening caused by the strain field formed by the mismatch between the Cr and Fe atoms [64-66].

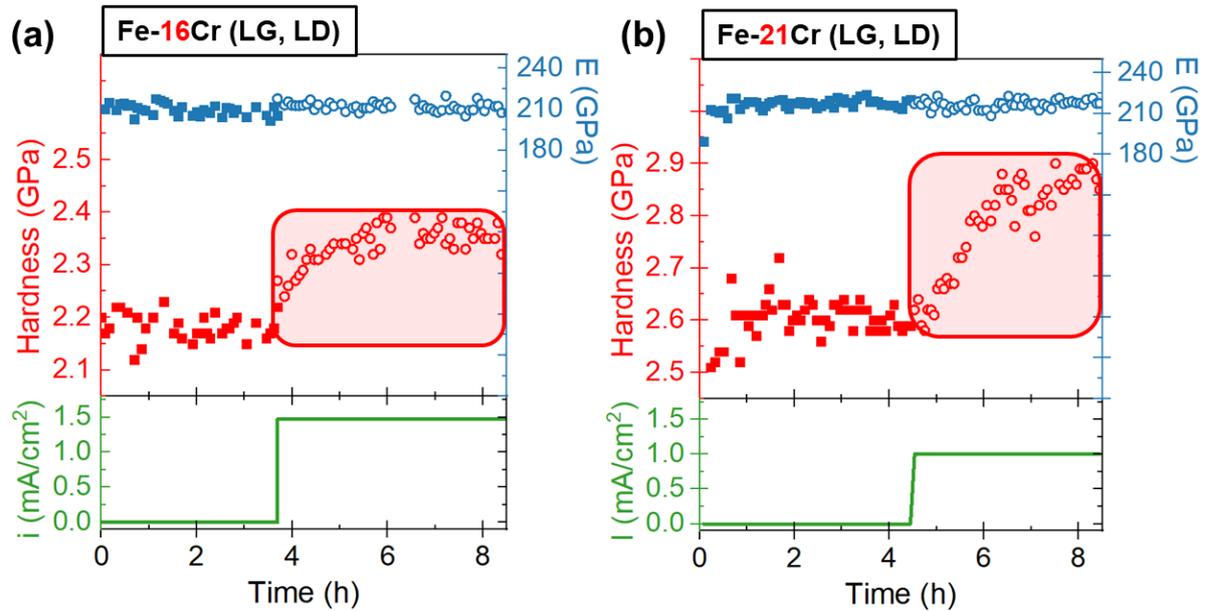

Fig. 8. Young's modulus (blue) and hardness (red) evolution with the current density (green) by applying a potential of -1.25 $V_{Ref}$ in (a) Fe-16Cr (LG, LD) and (b) Fe-21Cr (LG, LD) after hydrogen pre-charge and release of lightly trapped hydrogen and diffusive hydrogen.

At the onset of a constant hydrogen charging, the hardness increases and follows a sigmoidal shape until reaching a plateau where the quasi-equilibrium state of hydrogen flux is attained (red hollow circles). The Young's modulus (blue marks) remains unaltered before and during the hydrogen charging as the setup is stable. The previous TDS analysis shows that Fe-21Cr absorbs more hydrogen of 0.58±0.01 wt.ppm than Fe-16Cr (0.44±0.02 at.ppm). From the perspective of Zhang et al. [67], these hydrogen solubilities are low to cause any changes in the Young's modulus. With a higher hydrogen content, Fe-21Cr exhibits a more pronounced hardness enhancement of 0.24 GPa during hydrogen charging compared to Fe-16Cr of 0.18

GPa with the same applied potential of -1.25 $V_{Ref}$ (Fig. 8). This hydrogen-induced hardening effect occurs in α-Fe and 13Cr binary alloy as well [68].

*3.5. Comparison between the nanohardness-based and KP-based hydrogen permeation behavior (Effective diffusion coefficient)*

During the collection of the nanohardness evolution curve by employing the *in situ* backside nanoindentation hydrogen charging setup, a continuous Argon flow is purged onto the sample surface to prevent the release of hydrogen and water formation. The hydrogen concentration in the alloy on the frontside (exit or testing side) surface could be considered 0 ($C_2$) throughout the experiments. In addition, the backside (entry or charging side) of the specimen is under continuous electrochemical hydrogen charging supply, the hydrogen concentration upon which is treated as a constant value of $C_0$ ($C_1=C_0$). The initial hydrogen content is 0 throughout the bulk material ($f(x')=0$). Based on the Fick's law, the whole system can be interpreted by the non-steady state diffusion process with constant surface concentrations in a plane sheet model [69]. In summary, the boundary conditions during hydrogen charging for the system are elucidated by the following equations:

$$C_1=C_0, x=0, t\geq 0 \quad (4)$$
$$C_2=0, x=L, t\geq 0 \quad (5)$$
$$f(x)=0, 0<x<L, t=0 \quad (6)$$

where *L* is the sample thickness, the solution for the accumulation of hydrogen diffusion flux (*J*) at the side with negligible hydrogen concentration (*x=L*) obtained by the Laplace transform or the separation of the variable is given as [70]:

$$\frac{J(x)}{J_\infty}=1+2\sum_{1}^{\infty} \cos\frac{n\pi x}{L}\exp\left(-\frac{Dn^2\pi^2 t}{L^2}\right) \quad (7)$$

where $J_\infty$ represents the condition when the exit hydrogen flux reaches the steady state.

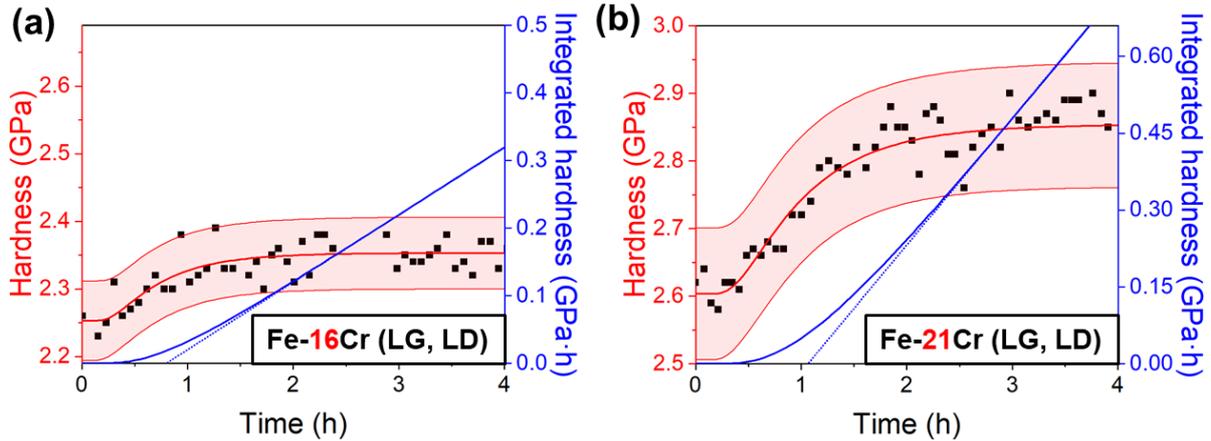

Fig. 9. The hardness evolution after hydrogen pre-charging at the commencing of hydrogen supply (black square dots) is simulated based on the non-steady state model of Eq. (7) (solid red line) and the integration of the hardness (solid blue line), and the linear fit of the integrated hardness (dotted blue line) for (a) Fe-16Cr (LG, LD) and (b) Fe-21Cr (LG, LD). The light red region indicates the 95 % confidence band for the simulated hardness curve.

The black squares in Fig. 9 are the original hardness data for Fe-16Cr (LG, LD) and Fe-21Cr (LG, LD) upon charging of the sample, taken from the red square highlighted region of Fig. 8. The proportional relationship between the hydrogen flux and the hardness values measured close to the sample surface can be supported by the observed linear increase in nanohardness with enhanced applied current density. This was shown for similar materials, Fe-16Cr (LG, LD) and Fe-21Cr (LG, LD), during *in situ* backside hydrogen charging [71]. Besides, Dafft et al. demonstrated a positive proportional relation between the applied cathodic current density and hydrogen activity at ambient temperature [72].

Therefore, as an indicator of the hydrogen flux on the sample surface, the change in the hardness values can be used to simulate the hydrogen permeation behaviour through the sample. In Fig. 9, the solid red lines are the simulated curves based on Eq. (7), considering the boundary conditions on the front and back sides of the specimen from Eq. (4-6). Due to the hydrogen concentration gradient across the specimen, the hydrogen flux continuously permeates through it, leading to a hardness enhancement in the exit side until reaching the steady state. The hydrogen flux stays constant as the absorbed and desorbed hydrogen reach a steady state. By integrating the hardness evolution curve, we can obtain the hydrogen permeation curve, which represents the accumulated hydrogen content as a function of time as shown in Fig. 5b. Hence, by linear fitting this integrated hardness evolution curve, the $t_{lag}$ is determined on the x-axis for further hydrogen diffusion coefficient calculation.

Since the permeation behavior of these two alloys is independent of the input hydrogen concentration based on this model, it is possible to compare the diffusion coefficients for these two alloys charged with different current densities [73]. Only the $J_\infty$ value alters with various input hydrogen concentrations. A summation of $n$=16 has been applied in simulating the hardness evolution, which is sufficiently accurate with n>6 according to the ASTM G148-97(2018) standard [74].

Table 4    Hydrogen effective diffusion coefficient calculated from the KP-based potentiometric hydrogen electrode method and the nanohardness-based method in ferritic Fe-16Cr (LG, LD) and Fe-21Cr (LG, LD) alloys.

| Method | Fe-16Cr (LG, LD) | Fe-21Cr (LG, LD) | $D_{eff(Fe\text{-}16Cr)}/ D_{eff(Fe\text{-}21Cr)}$ |
|---|---|---|---|
| KP-based $D_{eff}$ (cm$^2$/s) | (5.29±0.10)×10$^{-6}$ | (3.64±0.03)×10$^{-6}$ | 1.5±0.1 |
| Nanohardness-based $D_{eff}$ (cm$^2$/s) | (3.1±0.6)×10$^{-6}$ | (1.8±0.3)×10$^{-6}$ | 1.7±0.3 |

### 4. Discussion

*4.1. Differences of $D_{eff}$ between nanohardness-based and KP-based hydrogen permeation methods*

The effective diffusion coefficients obtained for Fe-16Cr (LG, LD) and Fe-21Cr (LG, LD) are listed in Table 4. The ratio of $D_{eff(Fe\text{-}16Cr)}$ and $D_{eff(Fe\text{-}21Cr)}$ using the method of KP-based is 1.5±0.1, similar to the ratio applying the nanohardness-based method of 1.7±0.3. Besides, the $D_{eff}$ values are within the same order of magnitude for both methods, confirming the feasibility of utilizing the nanohardness-based method to collect the hydrogen diffusion coefficient in ferritic alloys. However, it is noticeable that the $D_{eff}$ value obtained using the KP-based method is slightly larger than that obtained using the nanohardness-based method. This difference can be attributed to the discrepancy in the $C_2$ values between the two systems. In the KP-based method, an extra 100 nm Pd layer is deposited over the inherent Cr oxide layer of approximately 4 nm that naturally forms on the Fe-21Cr [71]. This Pd foil, which has a lower hydrogen chemical potential than the FeCr alloys [42, 43], absorbs hydrogen from the sample, causing a depletion of hydrogen near the outermost surface of the specimen. Combining with the properly sealed experimental chamber filled with nitrogen at controlled humidity of 0 rh %, the $C_2$ value remains consistently at zero throughout the experiment in the KP-based method.

However, in the nanohardness-based method, the hydrogen flux motion could be hindered by the intrinsic dense oxide layer. In addition, the nanoindentation chamber is purged with Argon through a pipe without humidity control. Consequently, we can assume that $C_2$ starts at zero and gradually increases to an equilibrium level upon reaching the break through time. Therefore, in the case of the nanohardness-based method, the hydrogen concentration gradient through the sample thickness becomes flatter as a function of time, leading to a lower hydrogen diffusion coefficient. Moreover, the nanohardness method acquires data approximately every 5 min, which is significantly longer compared to the KP-based method that collects data every 10 s. This larger time interval leads to a larger error for the nanohardness method. Apart from the factors mentioned above, the variation in the tip's diameter and electrolyte state used in both methods, may also contribute to the difference between the $D_{eff}$.

The nanohardness-based permeation method allows obtaining both, the mechanical and diffusion data simultaneously in bulk materials. It should be noted that we are not proposing a new method of measuring the diffusion coefficient. Instead, this method gives a new prospect in applying the novel *in situ* nanoindentation technique that can unveil more information from one series of experiments. Nevertheless, if the material hardness is not linearly related to the applied current density, for example when phase transitions or hydride formation occur during hydrogen supply, this nanohardness-based method is no longer applicable.

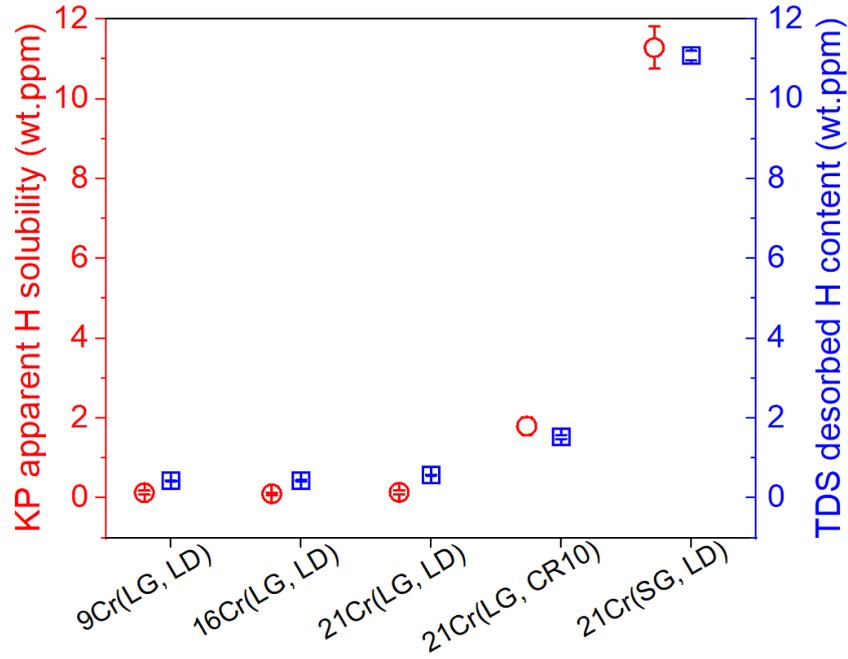

Fig. 10.    A comparison of the apparent hydrogen solubility acquired by using the KP-based permeation method and the desorbed hydrogen content obtained from TDS for the different alloys after hydrogen saturation by electrochemical charging at the applied current density of ~2 mA/cm$^2$.

### *4.2. Comparative analysis of hydrogen solubility by TDS and KP-based methods*

Based on the KP-based permeation experiments, the apparent hydrogen solubility ($C_{KP.\ H}$, wt. ppm) is composed of the apparent mobile hydrogen solubility ($C_{mo.\ H}$) and the apparent trapped hydrogen solubility ($C_{Trap.\ H}$), which can be calculated using the following equations [44]:

$$C_{mo.\ H} = \frac{q_{Hd,0} * A_s * M_H}{W_{steel}} * 10^6 \quad (8)$$

$$q_{Hd.\ 0} = \frac{i_p * L^2}{2 * F * D} \quad (9)$$

where $q_{Hd,0}$ is the molar amount of diffusible hydrogen under the steady-state of the permeation flux; $A_s$ is the surface area; $M_H$ is the molar mass of hydrogen; $W_{steel}$ is the weight of the specimen.

$$C_{Trap.\ H} = \frac{N_T * V * M_H}{W_{steel}} * 10^6 \quad (10)$$

$$N_T = \frac{3 \cdot i_p \cdot (t_{lag} - t_L)}{F \cdot L} \tag{11}$$

where $N_T$ is the molar density of all the traps in the specimen; V is the volume of the specimen; $t_L$ is the lag time of the lattice diffusion of hydrogen in pure FeCr alloy, which can be calculated based on Eq. (2). Here, we used the $D_{eff}$ values obtained above in this study as the lattice diffusion coefficients for the different alloys, accounting for the filling of deep hydrogen traps.

Fig. 10 shows the apparent hydrogen solubility ($C_{KP.\,H}$, in wt.ppm) measured by the KP-based permeation device and the desorbed hydrogen content captured by TDS ($C_{TDS.\,H}$, in wt.ppm) and the corresponding data is listed in Table 2. The difference in sample thickness in these two types of measurements (Supplemental Material Table S1) is considered the reason behind the lower measured $C_{KP.\,H}$ value compared to $C_{TDS.\,H}$. The sample thickness of Fe-Cr alloys with different Cr content used in the KP permeation tests is ~2.4 times of that used for TDS investigations. Based on the findings in [44], the sample thickness has an inverse linear relationship with $i_p$, which is critical for computing the apparent hydrogen solubility according to Eqs. (8-11). Nevertheless, the same tendency in hydrogen solubility can be noticed: a higher hydrogen content is present in alloys with higher Cr content, higher dislocation density and smaller grain size.

*4.3. Different hydrogen trapping sites*

The hydrogen desorption rate curves shown in Fig. 4a are mainly composed of three peaks of 79±10 °C, 136±18 °C and 435±19 °C as investigated by multi-peak fitting method (Fig. S1). The peaks values are an average for all the 5 alloys (details of the peak position can be found in supplement materials Table. S2). The small amount of C and O remaining in the alloys (Table 1) contribute to the formation of carbides (~40 x 200 $nm^2$) distributed along the grain boundaries and Cr oxides (~4 μm) embedded in the matrix, according to our previous studies [71]. Carbides and Cr oxides that are assumed to be deep hydrogen trapping sites, as illustrated in [33], with binding energies of 65 kJ/mol and 51-70 kJ/mol, respectively. Desorbed hydrogen peaks at 435±19 °C have a small contribution (~0.0003 wt.ppm/s), as the amount of carbides and oxides is low in all the materials, compared with the peaks at lower temperatures <200 °C. The experimental evidence for Cr carbides acting as hydrogen trapping sites in ferritic steels was also proposed by Depover et al. by conducting TDS measurements [27].

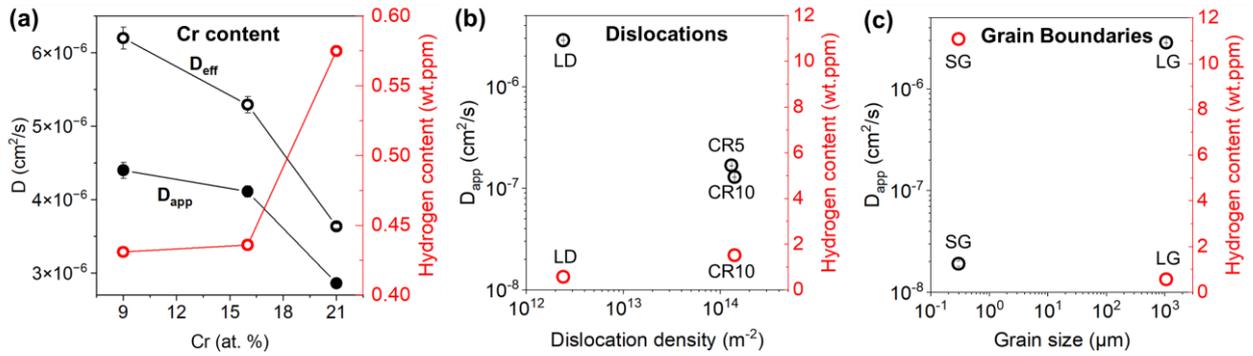

Fig. 11. (a) Comparison between $D_{app}$ and $D_{eff}$ in Fe-9Cr (MG, LD), Fe-16Cr (LG, LD) and Fe-21Cr (LG, LD) investigated by the KP-based potentiometric hydrogen permeation method. $D_{app}$ and hydrogen content measured by TDS compared between FeCr alloys with (b) different dislocation densities and (c) different grain sizes.

Fig. 11a shows the effective and apparent hydrogen diffusion coefficients in Fe-9Cr (MG, LD), Fe-16Cr (LG, LD) and Fe-21Cr (LG, LD) measured by the KP-based potentiometric hydrogen permeation method. Without the influence of deep hydrogen traps, $D_{eff}$ decreases with increased Cr content, following almost a linear relationship. The reduction of the hydrogen diffusion coefficient of 200 times in Fe-9 wt.%Cr compared with α-Fe was also found by Ramunni et al. [37] based on numerical calculations. Furthermore, TDS measurements (Fig. 4a and Table 2) showed that more hydrogen is absorbed in the alloys with higher Cr content, which is consistent with hot extraction chemical experiments and ab-initio calculations in [3, 37]. The only difference between Fe-16Cr and Fe-21Cr is their Cr content since the experiments were performed in materials subjected to long-term annealing treatments, reaching similar grain sizes, dislocation densities and vacancies. Hence, it is reasonable to assume that Cr, as the principal substitutional element, is responsible for increasing the hydrogen intake and reducing hydrogen diffusivity in FeCr alloys. The crystalline misfit generated by the size difference between Cr and Fe atoms causes a lattice distortion, creating more stable octahedral interstitial sites leading to enhanced hydrogen occupation as proposed in [37]. This normalized lattice dilatation caused by the presence of Cr in the binary FeCr alloys has been verified by X-ray diffraction [35]. In addition, considering the Pauling scale, hydrogen has a higher chemical affinity for Cr than for Fe. Finally, the Cr hydrogen binding energy in ferritic iron is 26-27 kJ/mol, according to the literature [33, 34]. Therefore, Cr atoms induce a volume change of the interstitial sites in FeCr alloy and create a new type of octahedral interstitial sites, acting in turn as reversible hydrogen trapping sites.

As shown in Fig. 11a, $D_{eff}$ is larger than $D_{app}$ for the same material, primarily because the deep traps that limit the motion of hydrogen are filled in advance. In addition, $D_{app}$ for Fe-16Cr is close to that for Fe-9Cr. This can be explained by the fact that grain boundaries are trapping sites for hydrogen, as aforementioned. Therefore, the higher amount of grain boundaries in Fe-9Cr would cause a stronger reduction in $D_{app}$ as expected for larger grains, as those in Fe-16Cr.

As shown in Fig. 6 and Table 3, $D_{app}$ has an intimate correlation with the hydrogen traps available in the samples. Note that the influence of hydrogen traps in the Pd coating and the Pd-sample interface are neglected in this study, as also verified for [44]. The Pd coating deposition parameters and the mechanical surface preparation is the same for all the alloys, therefore the hydrogen behavior in these traps is similar in all cases and their effect can be disregarded in this context. Therefore, the intrinsic traps in the materials play the predominant role in altering the hydrogen diffusion coefficient.

Fig. 11b shows that the amount of hydrogen absorbed in the cold-rolled Fe-21Cr (LG, CR10) increased drastically compared to Fe-21Cr (LG, LD), and the apparent diffusion coefficient decreased by one order of magnitude accordingly. The majority of the hydrogen in Fe-21Cr (LG, CR10), with a higher dislocation density, is desorbed at ~79 °C, as opposite to Fe-21Cr (LG, LD) where the major peak occurs at instead of at ~136 °C (Fig. S1). This suggests that dislocations are responsible for the hydrogen released at ~79 °C. In similar experiments using single crystal α-Fe, TDS measurements after cathodic hydrogen charging with a current density of 2 mA/cm$^2$, showed a peak at 122 °C in the hydrogen evolution rate curve, which was attributed to the hydrogen released from dislocations with a heating rate of 1°C/min [15]. Typically, a peak broadening and shift to higher temperatures of the hydrogen desorption peak are observed by TDS measurements with increasing heating rates [75, 76]. Deuterium trapped at dislocations has been also resolved by cryogenic atom probe tomography [30]. After the cold rolling treatment, the dislocation density in Fe-21Cr increased from $(2\pm2)\times10^{12}$ m$^{-2}$ to cold-rolled of $(1.4\pm0.1)\times10^{14}$ m$^{-2}$, while the grain size remained within 10 % alteration, as illustrated in Table 3. As a result, it can be concluded that the dislocations present in this alloy act as reversible hydrogen trapping sites, since also deeper hydrogen traps do not have an impact on the permeation test.

The HPT-treated Fe-21Cr (SG, LD) has a smaller grain size, in the submicrometer range, and a lower dislocation density when compared to Fe-21Cr (LG, CR10). As shown in the EBSD analysis of Fig. S1, Fe-21Cr (SG, LD) contains 86 % high angle grain boundaries with larger than 15° rotation angles. Fig. 11b exhibits a reduction of ~7 times reduction in $D_{app}$ that can be attributed to the grain boundary trapping effect, as the lower dislocation density enhances the diffusion coefficient based on our previous discussion (Table 3). Furthermore, the absorbed hydrogen amount is also ~7 times higher in Fe-21Cr (SG, LD) than in Fe-21Cr (LG, CR10). This reduced grain size, leads to a larger contribution of the ~136 °C peak in Fe-21Cr (SG, LD) compared with Fe-21Cr (LG, LD). It is then reasonable to assume that grain boundaries are responsible for the peak at ~136 °C. This result is consistent with the desorbed hydrogen at a heating rate of 1 °C/min in polycrystalline iron, where the hydrogen peak at 142 °C is assigned to grain boundary traps [15]. As interference in the permeation experiment from other deep hydrogen traps has been avoided by modifying only the grain size, grain boundaries are proposed to be the reversible trapping sites that decrease $D_{app}$ in these FeCr alloys. This coincides with kinetic Monte Carlo studies of Du et al. [12, 13] in bcc Fe.

In summary, interstitial sites introduced by dislocations, grain boundaries and Cr atoms are proposed in this study to be the three kinds of reversible hydrogen trapping sites that influence the first peaks between 50-200 °C in these ferritic FeCr alloys for TDS measurements at the specific heating rate of 16 °C/min. This assumption coincides well with the study of Shi et al., which also identified the peak at ~150 °C in α-Fe as the hydrogen absorbed by dislocations as reversible hydrogen traps with the same TDS technique but a heating rate of 5 °C/min [75]. With a heating rate of ~25 °C/min in a bcc Fe-5 wt.% Ni model alloy, the two dominant peaks observed after cold rolling and hydrogen charging for 2 h were located at 100-200 °C and 200-300 °C. These two peaks were identified as dislocations (desorption energy 29 ± 5 kJ/mol) and vacancies (desorption energy 38 ± 5 kJ/mol), respectively [31]. In a low carbon ferritic steel, cathodically charged with hydrogen, Chen et al. also assigned the 100-200 °C TDS peak to the hydrogen trapped by dislocations/grain boundaries [30].

## 5. Summary and Conclusions

We have performed a novel efficient nanohardness-based permeation experiment using an *in situ* backside nanoindentation setup, which enables us to measure the mechanical properties and the diffusion coefficient of the bulk FeCr alloys simultaneously. In addition, the KP-based potentiometric hydrogen electrode method and TDS analysis were used to investigate hydrogen interactions with various trapping sites in ferritic FeCr alloys, considering different chromium content, dislocation densities, and grain sizes. The measured hydrogen diffusion coefficients are classified into apparent and effective diffusion coefficients depending on whether deep hydrogen traps are involved or not. The major conclusions are as follows:

1) The nanohardness-based effective diffusion coefficient obtained from the *in situ* backside nanoindentation setup is $(3.1\pm0.6)\times10^{-6}$ cm$^2$/s for Fe-16Cr (LG, LD), which is higher than for Fe-21Cr (LG, LD) with higher Cr content, of $(1.8\pm0.3)\times10^{-6}$ cm$^2$/s. These effective hydrogen diffusion coefficients are comparable to those obtained from the KP-based potentiometric hydrogen electrode method, which are $(5.98\pm0.11)\times10^{-6}$ cm$^2$/s for Fe-16Cr (LG, LD) and $(3.64\pm0.03)\times10^{-6}$ cm$^2$/s for Fe-21Cr (LG, LD). This comparison verifies the viability of this apparatus in simultaneously measuring the mechanical properties and hydrogen diffusion coefficient, and opens the possibility to use the nanohardness data as hydrogen markers in further experiments in other alloys.

2) Hydrogen trapping at dislocations was analyzed in Fe-21Cr with similar grain size, but two orders of magnitude different dislocation densities from $(2\pm2)\times10^{12}$ m$^{-2}$ to $(1.4\pm0.1)\times10^{14}$ m$^{-2}$, measured from ECCI. This increase in dislocation density results in an enhancement of the absorbed hydrogen amount from 0.58±0.01 wt.ppm to 1.53±0.05 wt.ppm and a reduction of apparent hydrogen diffusion coefficient from $(2.86\pm0.03)\times10^{-6}$ cm$^2$/s to $(1.29\pm0.01)\times10^{-7}$ cm$^2$/s for Fe-21Cr (LG, LD) and Fe-21Cr (LG, CR10), respectively.

3) The impact of grain boundaries on the hydrogen diffusion and uptake was analyzed by decreasing the grain size from 1049±51 µm to 0.3±0.1 µm. This reduction in the grain size resulted in a lower apparent hydrogen diffusion coefficient of (1.92±0.12)×$10^{-8}$ cm$^2$/s when compared to Fe-21Cr alloy with larger grain size, (2.86±0.03)×$10^{-6}$ cm$^2$/s. A corresponding increase of absorbed hydrogen for the smaller grain size alloy was also observed, from 0.58±0.01 wt.ppm to 11.08±0.12 wt.ppm.

4) The apparent hydrogen solubility measured by the KP-based permeation method and the desorbed hydrogen content captured using TDS show consistency in the tendency of hydrogen being trapped in different materials. Dislocations, grain boundaries and Cr interstitials are found to be the reversible trapping sites for hydrogen that drive a decrease in the diffusion behavior and increase the amount of strapped hydrogen.

The insights gained from studying the interactions between hydrogen and the microstructure features studied here (i.e. dislocations, grain boundaries and Cr interstitials) can be extended to other iron-based ferritic steels that share similar structures. Consequently, this perspective provides valuable guidance in the design of hydrogen-resistant materials by either incorporating or avoiding the formation of various atomic-scale hydrogen traps. The approach of applying the backside *in situ* nanoindentation for studying HE in Fe-Cr alloys can also be further allocated to other iron-based ferritic materials. With further modifications, like sealing the whole system or modifying the specimens' dimensions, its usage range can be even extended to other alloy systems with different FCC or HCP structures and coated materials [77].

**Declaration of Competing Interest**

The authors declare that they have no known competing financial interests or personal relationships that could have influenced the work reported in this paper.

**Acknowledgements**

The project was funded by the German research foundation (Deutsche Forschungsgemeinschaft, DFG) in project number 318876084 "hydrogen-microstructure interactions in ferritic alloys at small scale". Gerhard Dehm and María Jazmin Duarte acknowledge financial support by the ERC grant GB correlate (787446). Jing Rao acknowledges financial support by the KSB foundation.

## 6. Supplementary material

*6.1. EBSD mapping of Fe-21Cr (SG, LD)*

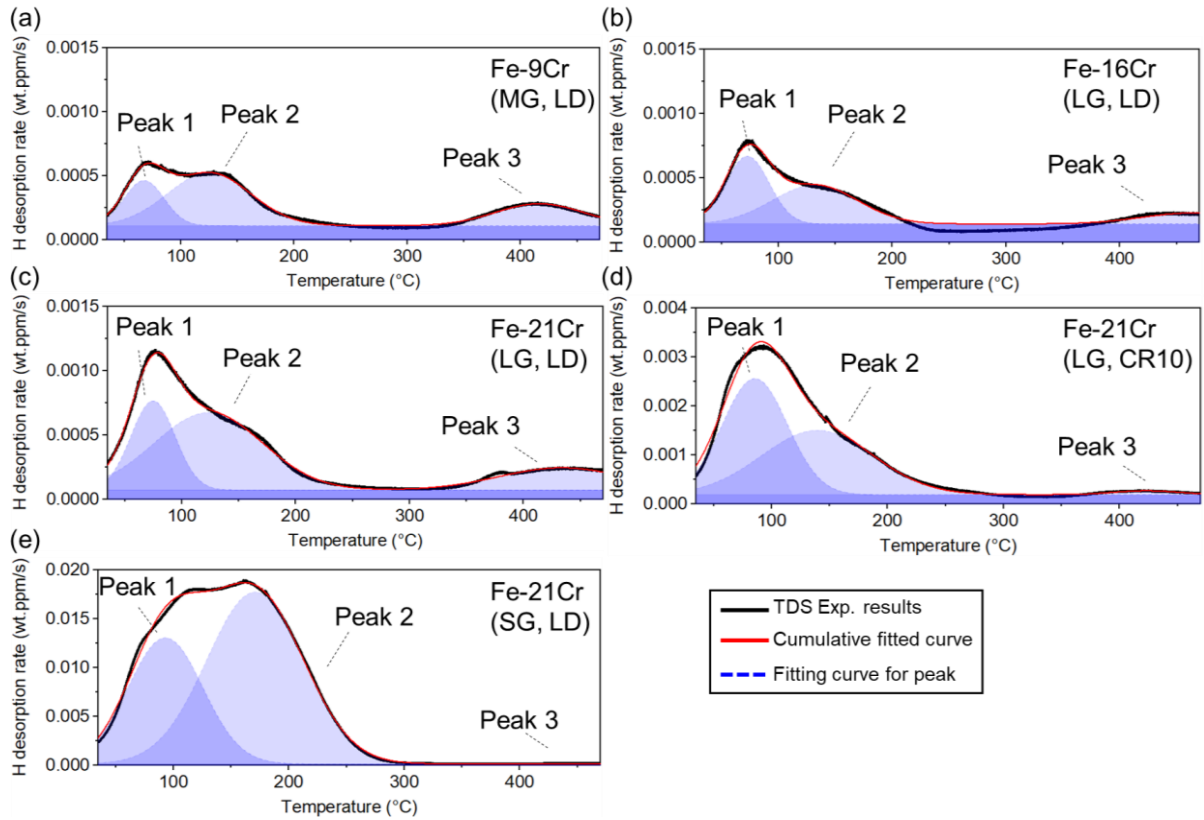

Fig. S1. Multi-peak fitting of the hydrogen desorption curves (Fig. 4a) based on the Gaussian peak function. Note the difference in scales for the coordinate axis.

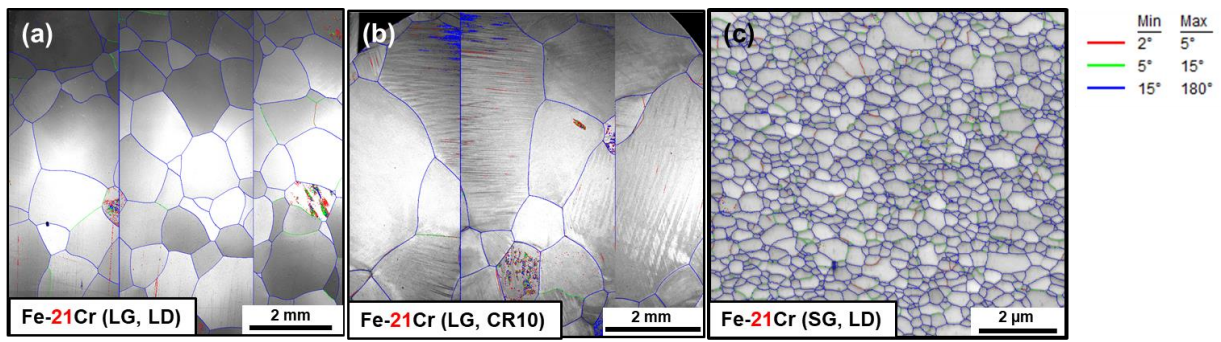

Fig. S2. Grain boundary mapping of Fe-21Cr (LG, LD), Fe-21Cr (LG, CR10) and, Fe-21Cr (SG, LD) investigated by the EBSD.

Table S1 Thickness of the sample used in KP and TDS measurements (in mm).

|  | Fe-9Cr MG, LD | Fe-16Cr LG, LD | Fe-21Cr LG, LD | Fe-21Cr LG, CR10 | Fe-21Cr SG, LD |
|---|---|---|---|---|---|
| KP | 2.22±0.02 | 2.21±0.02 | 2.23±0.01 | 2.44±0.03 | 0.87±0.02 |
| TDS | 1.0±0.1 | 1.0±0.1 | 1.0±0.1 | 1.0±0.1 | 0.9±0.1 |

Table S2 Peak position for Fig. S1 (in °C).

|  | Peak 1 | Peak 2 | Peak 3 |
|---|---|---|---|
| Fe-9Cr MG, LD | 67.4±0.1 | 124.8±0.1 | 416.7±0.1 |
| Fe-16Cr LG, LD | 72.6±0.1 | 131.8±0.6 | 464.7±0.6 |
| Fe-21Cr LG, LD | 85.5±0.1 | 139.3±1.1 | 423.3±0.8 |
| Fe-21Cr LG, CR10 | 74.6±0.1 | 123.3±0.3 | 433.5±0.3 |
| Fe-21Cr SG, LD | 93.0±0.1 | 170.7±0.1 | - |
| Average | 79±10 | 136±18 | 435±19 |